\documentstyle[preprint,aps]{revtex}

\begin{document}

\title{Molecular nanomagnets in ac magnetic field.}

\author {N.A. Sinitsyn $^1$}
\address{$^1$Department of Physics, Texas A\&M University, College Station, 
Texas 77843-4242\\
e-mail: nsinitsyn@physics.tamu.edu}

\maketitle
\date{ } 
\begin{abstract}
  The behavior of molecular nanomagnets in periodic magnetic field transverse to the easy axis direction is investigated.
  It is shown that at sufficiently
  strong field the tunneling time can be considerably reduced.  
\end{abstract}
\newpage


Recent years there was a great growth of interest in molecular nanomagnets. It was proven experimentally
that they display quantum hysteresis and quantum tunneling at the molecular level  \cite{Th}, \cite{Fr}, \cite{Wern2}, \cite{tunneling},
nonexponential dynamic scaling of the magnetization relaxation \cite{nonexp} and
tunneling rate variations due to Berry phase \cite{Wern}, \cite{Berry}, \cite{Berry2}.
Many efforts have been made to observe the quantum coherence \cite{coh}. The known nanomagnets possess a
strong magnetic anisotropy. At sufficiently low temperature they can be described by a single spin Hamiltonian.
 The energy eigenvalues of such Hamiltonian are approximately doubly degenerate. The degeneracy is lifted by the tunneling caused by 
transverse anisotropy. Generally, the level splitting is very small but it can be enhanced sufficiently by a transverse to the easy axes magnetic
field \cite{transv}, \cite{transv1}.
At very low temperature only the tunneling between the states of the ground doublet are important.
As was demonstrated in \cite{Vug} additional possibility to enhance tunneling rate is to apply ac magnetic field with the resonance frequency. 
  
Suppose that if $B_1=0$ then $|0_s>,|0_a>$ and $|1_s>,|1_a>$ are accordingly the lowest symmetric and antisymmetric and first excited 
symmetric and antisymmetric eigenstates of the
Hamiltonian, respectively,
\begin{equation}
\hat{H}_1=-D\hat{S}_z^2-g\mu_B B_0\hat{S}_x+g\mu_B B_1\hat{S}_x \cos(\omega t)
\label{h1}
\end{equation}
or
\begin{equation}
\hat{H}_2=-D\hat{S}_z^2-E\hat{S}_x^2+g\mu_B B_1\hat{S}_x \cos(\omega t)
\label{h2}
\end{equation}

Without ac magnetic field, system can be truncated to the lowest two level one with tunneling rate $\Delta_0$.
Let denote the energy difference of the 
first and second symmetric levels as $\omega_0$.  The gap $\Delta_1$ between states $|1_s>$ and $|1_a>$ is much bigger than the gap $\Delta_0$ 
between $|0_s>$ and $|0_a>$.
The idea of the work \cite{Vug} is that, if the frequency $\omega$ is exactly equal to $\omega_0$ and the linewidth of the ac field is much smaller than
$\Delta_1$, the whole
 system
can be truncated to an effective three level one. As a result the following time dependence of magnetization was found
\begin{equation}
<\hat{S}_z(t)>=S\cos(\Omega_Rt)\cos(\Delta_0t)
\label{vf}
\end{equation}
here $\Omega_R$ is the Rabi frequency.
 If the amplitude $B_1$ of the ac field is strong enough, the Rabi frequency $\Omega_R$ can be much higher than $\Delta_0$ and
  magnetization oscillates much faster. 
 Nevertheless this result is applicable only if $g\mu_B B_1 \sqrt{S}<<\Delta_1$, otherwise, as we show further the state $|1_a>$
  becomes also important. 
 Usually the ratio $\frac{\Delta_0}{\Delta_1}$ is of the order $10^{-2}-10^{-3}$.
  This means that one can enhance tunneling rate not more than $10-100$ times.
 
 In this article we propose to apply much stronger ac field in the range $\Delta_1<<g\mu_B B_1\sqrt{S}<<\omega_0$. The level $|1_a>$ becomes important
 and system can be truncated already to the effective four level one.

Suppose that $\omega$ is close to resonance frequency $\omega_0$ so that Rotating Wave Approximation is applicable. 

At any moment the vector of state can be written in the following form:
\begin{equation}
|\psi (t)>=a_1(t)|0_s>+a_2(t)|0_a>+|b_1(t)|1_s>+b_2(t)|1_a>
\label{vec1}
\end{equation}
If we denote 
\begin{equation}
h=g\mu_B B_1<1_s|\hat{S}_x|0_s>=g\mu_B B_1<1_a|\hat{S}_x|0_a>=g\mu_B B_1\sqrt{S/2}
\label{zeeh}
\end{equation}
The Shrodinger equation for the amplitudes $a_{s,a}, b_{s,a}$ reads:

\begin{equation}
-i\frac{d}{dt}a_1=h\cos(\omega t)b_1 
\label{eq1}
\end{equation}
$$
-i\frac{d}{dt}b_1=\omega_0 b_1+hcos(\omega t)a_1
$$
 
\begin{equation}
-i\frac{d}{dt}a_2=h\cos(\omega t)b_2 
\label{eq2}
\end{equation}
$$
-i\frac{d}{dt}b_2=(\omega_0+\Delta_2) b_2+hcos(\omega t)a_2
$$
Here we disregarded the value $\Delta_0$ since generally $\Delta_1>>\Delta_0$.

Changing the variables $b_{s,a}\rightarrow b_{s,a}e^{-i\omega t}$ and using the Rotating Wave Approximation 
we transform equations (\ref{eq1}) and (\ref{eq2})
in a following systems:

\begin{equation}
-i\frac{d}{dt}a_1=hb_1 
\label{eq3}
\end{equation}
$$
-i\frac{d}{dt}b_1=(\omega_0-\omega) b_1+ha_1
$$
and 
\begin{equation}
-i\frac{d}{dt}a_2=hb_2 
\label{eq4}
\end{equation}
$$
-i\frac{d}{dt}b_2=(\omega_0-\omega+\Delta_1) b_2+ha_2
$$

Solution of (\ref{eq3}) is

\begin{equation}
a_1(t)=e^{-i(\omega_0-\omega)t/2}(\cos(\Omega_R t)+\frac{i(\omega_0-\omega)}{2\Omega_R}\sin(\Omega_Rt))
\label{sol}
\end{equation}
$$
b_1(t)=e^{-i(\omega_0-\omega)t/2}\frac{-ih}{\Omega_R }\sin(\Omega_R t)
$$
where $\Omega_R=\sqrt{h^2+({\omega}_0-\omega)^2/4}$ is the Rabi frequency. Let the initial state was polarized up:
\begin{equation}
|\psi (0)>=\frac{1}{\sqrt{2}}(|0_s>+|0_a>)
\label{init}
\end{equation}
In the case $h>>\Delta_1$ the vector of state evolves with time as 
\begin{equation}
|\psi (t)>=\frac{1}{\sqrt{2}}(e^{-i\omega_st/2}\cos(ht)|0_s>+e^{-i\omega_a t/2}\cos(ht))|0_a>+
\label{vec2}
\end{equation}
$$
+ie^{-i\omega_s t/2-i\omega t}\sin(ht)|1_s>+ie^{-\omega_a t/2-i\omega t}\sin(ht)|1_a>
$$
where we denote $\omega_s=\omega_0-\omega$ and $\omega_a=\omega_0-\omega+\Delta_1$
The only nonzero elements of $\hat{S}_z$ are:
\begin{equation}
<0_s|\hat{S}_z|0_a>=S
\label{nz1}
\end{equation}
\begin{equation}
<1_s|\hat{S}_z|1_a>=S-1
\label{nz2}
\end{equation}
Hence
\begin{equation}
<\hat{S}_z(t)>=<\psi (t)|\hat{S}_z |\psi(t)>=S\cos(\Delta_1 t/2)\cos^2(ht)+(S-1)\cos(\Delta_1 t/2)\sin^2(ht)
\label{vec3}
\end{equation}
Averaging over fast oscillations with the frequency $h$, we find slow variation of magnetization:
\begin{equation}
<<\hat{S}_z(t)>>=S(1-\frac{1}{2S})\cos(\Delta_1 t/2)
\label{av}
\end{equation}

Since the ratio $\frac{\Delta_0}{\Delta_1}$ is typically of the order $10^{-2} - 10^{-3}$, the tunneling
rate can increase considerably. It is interesting that (\ref{av}) does not depend on $\omega$ so it may not exactly match in resonance. 
The only requirement is that rotating wave approximation should be valid. This makes
result less dependent on the quality of periodic magnetic field in contrast to the result \cite{Vug} where strong restrictions on $\Delta \omega$ were
important.

Besides possible application for macroscopic quantum coherence, result (\ref{av}) can be useful for studying relaxation processes in molecular 
nanomagnets. Interaction with environment produces a strong bias that prevents tunneling for most molecules but those that accidentally match in resonance.
Tunneling of these molecules changes surrounding magnetic field and other spins appear in resonance. This leads to nonexponential
 relaxation of magnetization.
The described picture is very rough and complete theory of this process is still not built. Our results predict strong increase of relaxation rate
in ac field near resonance. If the number of systems in resonance is proportional to  the gap between the lowest two levels then in resonant ac field
this gap is effectively $\Delta_1$ rather than $\Delta_0$. Tunneling rate for these nanomagnets also is $\Delta_1$ rather than $\Delta_0$. So totally
relaxation rate may increase by the factor $\frac{\Delta_1^2}{\Delta_0^2}$. So measurements of relaxation rate 
in time periodic magnetic field can provide more information about relaxation mechanisms.

\noindent {\it Acknowledgments}. This work was supported by NSF under the grant DMR 0072115 and by DOE under the grant DE-FG03-96ER45598.
 The author thanks V.L. Pokrovsky for discussion and important remarks.


\begin{thebibliography}{999}
\bibitem{Th} L. Thomas, F. Lionti, R. Ballow, D. Gatteschi, R. Sessoli, B. Barbara, Nature, 383, 145 (1996)
\bibitem{Fr} J.R. Friedman, M.P. Sarachik, J. Tejada, R. Ziolo, Phys. Rev. Lett. 76, 3830 (1996) 
\bibitem{Wern2} W. Wernsdorfer, , T. Ohm, C. Sangregorio, R. Sessoli, D. Mailly, C. Paulsen, Phys. Rev. Lett. 82, 3903 (1999)
\bibitem{tunneling} Quantum Tunneling of Magnetization, L. Gunther, B. Barbara (Eds.), Kluwer Academic Publishers, Dordrecht, 1995
\bibitem{nonexp} L. Thomas, A. Caneschi, B. Barbara, Phys. Rev. Lett. 83, 2398 (1999)
\bibitem{Wern} W. Wernsdorfer, R. Sessoli, Science 284, 145 (1999) 
\bibitem{Berry} Ersin Kececiglu and A. Garg, cond-mat/0003319
\bibitem{Berry2} A.Garg, Europhys. Lett. 50, 382 (2000), cond-mat/0003113
\bibitem{coh} D.D. Awshalom, J.F. Smyth, G. Grinstein, D.P. DiVincenzo, D. Loss, Phys. Rev. Lett. 68, 3092 (1992)
\bibitem{transv} F. Lionti, L. Thomas, R. Ballou, B. Barbara, A. Sulpice, R. Sessoli, D. Gatteschi, J. Appl. Phys. 81, 4608 (1997)
\bibitem{transv1} G. Bellessa, N. Vernier, B. Barbara, D. Gatteschi, Phys. Rev. Lett. 83, 416 (1999)
\bibitem{Vug} I.D. Tokman, G.A. Vugalter, J. of Magn.Magn.Materials 222, 375-378 (2000) 

\end{thebibliography}
\end{document}